\documentclass[prl,twocolumn,showpacs,preprintnumbers,amsmath,amssymb]{revtex4}

\usepackage{graphicx}
\usepackage{amsmath}
\usepackage{dcolumn}
\usepackage{bm}

\begin{document}


\title{Reversible Transition Between Thermodynamically Stable Phases with Low Density of Oxygen Vacancies on SrTiO$_3$(110) Surface}

\author{Fengmiao Li$^1$, Zhiming Wang$^1$, Sheng Meng$^1$, Yongbao Sun$^2$, Jinlong Yang$^2$, Qinlin Guo$^1$, and Jiandong Guo$^1$}
\email{jdguo@aphy.iphy.ac.cn}.

 \affiliation{$^1$Beijing National Laboratory for Condensed-Matter Physics \& Institute of Physics, Chinese Academy of Sciences, Beijing 100190, China}
 \affiliation{$^2$Hefei National Laboratory for Physical Science at the Microscale \& University of Science and Technology of China, Hefei, Anhui 230026, China}

\date{\today}

\begin{abstract}
The surface reconstruction of SrTiO$_3$(110) is studied with scanning tunneling microscopy and density functional theory (DFT) calculations. The reversible phase transition between (4$\times$1) and (5$\times$1) is controlled by adjusting the surface metal concentration [Sr] or [Ti]. Resolving the atomic structures of the surface, DFT calculations verify that the phase stability changes upon the chemical potential of Sr or Ti. Particularly, the density of oxygen vacancies is low on the thermodynamically stabilized SrTiO$_3$(110) surface.
\end{abstract}

\pacs{68.47.Gh, 31.15.A-, 68.37.Ef, 68.35.Fx}
\keywords{oxide surfaces, Ab initio calculation, scanning tunneling microscopy, diffusion in interface}

\maketitle

Perovskite oxides have attracted intensive interests in the fields of fundamental condensed matter physics, photocatalysis chemistry, material science, as well as electronics applications, due to their rich phase diagrams and remarkable functionalities. The related research becomes even more exciting since the recent discovery of a quasi-two-dimensional electron gas (2DEG) between two insulating materials, SrTiO$_3$ and LaAlO$_3$ \cite{A. Ohtomo}. Subsequently, a tremendous amount of evidence has shown that the perovskite oxides in the low-dimensional (LD) form such as interfaces, thin films, or heterostructures display an equally rich diversity of exotic phenomena that is related, but not identical to the bulk \cite{JPCMRev}, indicating a great opportunity for novel oxide-based devices \cite{Mannhart2010}. One of the most important and intriguing discoveries is that the atomic arrangement on the surface or at the interface is determinant for the properties of the entire artificial structure. Hwang \textit{et al.} found that the formation of 2DEG critically depends on the type of atomic termination layer at the interface  \cite{A. Ohtomo}. O vacancies (V$_O$'s) also sensitively influence the density and mobility of the charge carriers at the heterointerface of LaAlO$_{3}$ and SrTiO$_{3}$ \cite{PRB07}. Therefore, to clarify the origin of the emergent properties in LD oxides and ultimately to tune them for the fabrication of functionalized devices, the detailed knowledge on their microscopic structures and the high-precision growth technique are the key issues.

Single crystalline SrTiO$_{3}$ is widely used as the epitaxial substrate for perovskite oxide films. In order to improve the growth quality, much effort has been made to obtain the atomically flat and ordered SrO or TiO$_2$ terminated (100) surface \cite{M. Kawasaki, G. Koster}. In contrast to the electrically neutral (100) surface, the formation process of SrTiO$_{3}$(110) surface structure is much more complicated since it is inherently unstable due to the perpendicular macroscopic dipole formed by alternatively stacked (SrTiO)$^{4+}$ and (O$_2$)$^{4-}$ layers \cite{C. Noguera}. The (110) surface stoichiometry often deviates from that in the ideal crystal, which leads to the formation of mixed phases of reconstruction \cite{SS97 Brunen}. V$_O$'s may also be responsible for the stabilization of the (110) surface \cite{APL05 Hwang}. Recently Russell \textit{et al.} obtained an (\textit{n}$\times$1) (\textit{n}=3,4,6) family of reconstructions at varying annealing temperatures, which was described as a homologous series with the TiO$_4$ tetrahedra model \cite{B.C. Russell, J.A. Enterkin}. However, it is still challenging to understand the stabilization mechanism of SrTiO$_3$(110) polar surface, particularly the behavior of oxygen (or vacancies).

In this letter, we investigate the microscopic structure of SrTiO$_{3}$(110) surface and present a way of preparing high-quality titanate surfaces and films with ideal oxygen stoichiometry. We obtain the monophased surface in (4$\times$1) or (5$\times$1) reconstruction. The reversible phase transition between them is realized by tuning Ti or Sr concentration on the surface. With \textit{ab initio} calculations, we determine the atomic structures of both phases and study their stability in different chemical environments. Thus the experimental control of surface reconstruction is understood within the thermodynamic picture. Moreover, both of the experimental observations and theoretical calculations reveal that the V$_O$'s density on SrTiO$_{3}$(110) surface could be extremely low.

The experiments were performed in an Omicron ultra-high vacuum (UHV) low temperature scanning tunneling microscopy (STM) system with the base pressure of 1$\times$10$^{-10}$ mbar. Nb-doped (0.7 wt$\%$) SrTiO$_3$(110) single crystals (12$\times$3 mm$^2$) were purchased from Hefei KMT Co., China. The as-received sample was sputtered with Ar$^+$ beam at room temperature followed by annealing in UHV. Strontium and titanium were evaporated from Knudsen cells, respectively. The flux of Sr was calibrated by observing the formation of Sr/Si(001)-(2$\times$1) reconstruction \cite{J.W. Reiner} with RHEED and STM, while the flux of Ti was calibrated by monitoring the RHEED intensity oscillation during the homoepitaxial growth of SrTiO$_3$(110) film \cite{Z. Wang}. The sample was resistively heated by passing a direct current and the temperature was monitored with an infrared pyrometer. DFT calculations were carried out with the \textquotedblleft Vienna \textit{ab initio} simulation package \textquotedblright (VASP) code \cite{G. Kresse}. We use the projector augmented-wave (PAW) method \cite{P.E. Blochl} and the Perdew-Burke-Ernzerhof functional \cite{J.P. Perdew} with the kinetic energy cutoff of 400 eV for plane waves and a (3$\times$5$\times$1) Monkhorst-Pack \textit{k}-point mesh. The surface structure was modeled with a supercell that was symmetrical along the [110] direction, consisted of a 13-layer slab separated by a vacuum layer of 12 \AA. Atoms in the centeral three layers were fixed and other atoms were allowed to relax until the force on each atom was less than 0.02 eV/\AA. Simulated STM images were generated by integrating the local density of states (LDOS) between Fermi level (E$_F$) and E$_F$+1.5 eV in Tersoff-Hamann approximation \cite{J. Tersoff} with the constant density method \cite{D.E.P. Vanpoucke}.

\begin{figure}
\includegraphics[clip,width=3.3in]{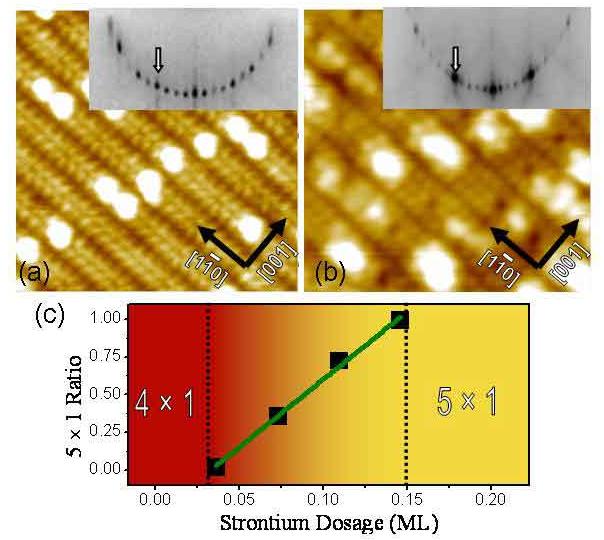}
\caption{(color online) (a) and (b) STM images (15$\times$15~nm$^2$, 1.5~V/20~pA) of (4$\times$1) and (5$\times$1) reconstructed SrTiO$_3$(110) surfaces, respectively. The insets show the RHEED patterns along [001] direction with (10) spots indicated by arrows, respectively. (c) The area ratio of (5$\times$1) domains through the entire surface depending on the Sr dosage. The statistics are done over 8 STM images (500$\times$500~nm$^2$) for each data point.}
\end{figure}

Figure~1~(a) shows the monophased (4$\times$1) reconstruction on the SrTiO$_3$(110) surface after Ar$^{+}$ sputtering (500~eV/2~$\mu$A for 10 minutes) followed by annealing at 1000$^o$C for 1 hour. The surface appears as periodic stripes along the [1$\overline{1}$0] direction, separated with a dark trench, and each contains two obvious bright rows of periodic dots. The periods along [1$\overline{1}$0] and [001] are $\sim$0.6~nm and 1.6~nm, respectively, corresponding to the (4$\times$1) reconstruction as observed clearly by RHEED. Evaporating a small amount of Sr metal followed by annealing at $1000\,^{\circ}\mathrm{C}$, domains with a new ordering form on the surface. As shown in Fig.~1~(b), it also appears as stripes but wider than that in the (4$\times$1) domain, each containing three bright rows of periodic dots along the [1$\overline{1}$0] direction. The periods along [1$\overline{1}$0]and [001] are $\sim$0.6~nm  and $\sim$1.95~nm, respectively, corresponding to (5$\times$1) reconstruction. The (5$\times$1) domain area enlarges with the Sr evaporation dosage linearly until a monophased surface formed with $\sim$0.15~ML Sr [relative to bulk-truncated SrTiO$_3$(110), 1~ML=4.64$\times$10$^{14}$~atoms/cm$^2$, see Fig.~1~(c)]. Such a phase transition can be reversed by evaporating $\sim$0.15~ML Ti onto the (5$\times$1) surface followed by annealing at 1000$^o$C \cite{Z. Wang}. 

\begin{figure}
\includegraphics[width=3.3in,clip]{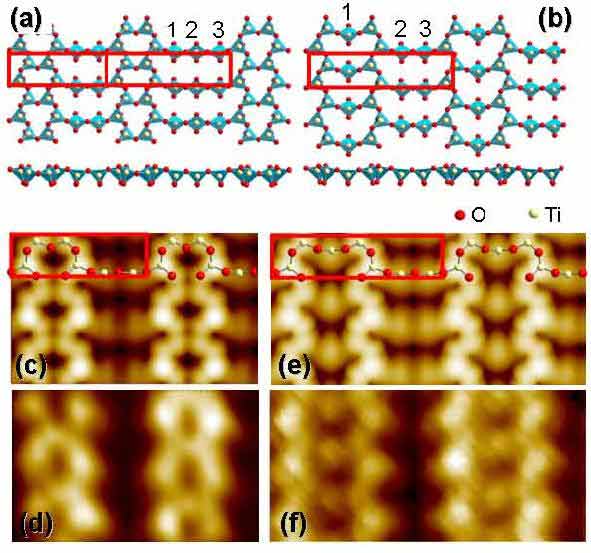}
\caption{(color online) (a) The structural model of the (\textit{n}$\times$1) reconstruction \cite{J.A. Enterkin} with (4$\times$1) on the left and (5$\times$1) on the right. (b) The modified structural model of (5$\times$1). The upper panels are the topview while the lower panels are the sideview. (c), (d) and (e), (f) The simulated and experimental STM images of (4$\times$1) and (5$\times$1), respectively.}
\end{figure}

With DFT calculations, different structural models of (4$\times$1) reconstruction are compared, including the Sr-adatom, TiO$_x$ and (100)-microfaceted models. We find that the TiO$_4$ tetrahedra model [the left part in Fig.~2~(a)] is energetically favorable, in consistence with what Enterkin \textit{et al.} reported previously \cite{J.A. Enterkin}. The simulated STM images reproduce the experimental features very well, as seen in Fig.~2~(c) and (d). The unoccupied states STM image shows both titanium and oxygen atoms on the surface in bright features, as indicated by the ball-and-stick drawing superimposed onto the simulated image. The zig-zag TiO$_4$ chains along [1$\overline{1}$0] correspond to the bright stripes in the STM images. 

Our experimental STM images of the (5$\times$1) reconstruction cannot be described by the (\textit{n}$\times$1) homologous series \cite{J.A. Enterkin} very well. We modify the initial structure by shifting TiO$_4$ tetrahedron 1 from the linear chain along [001] to the middle of the zig-zag chain along [1$\overline{1}$0], as shown in Fig.~2~(b). After relaxation, an extra energy gain of 0.6~eV is obtained comparing with the total energy calculated for the \textquotedblleft (\textit{n}$\times$1)\textquotedblright model \cite{J.A. Enterkin}. Although such a small difference could be the result of different functional used in calculations, the simulated image is in excellent agreement with the STM observation on the monophased (5$\times$1) obtained in the current work [Fig.~2~(e) and (f)]. 

\begin{figure}
\includegraphics[clip,width=3.4in]{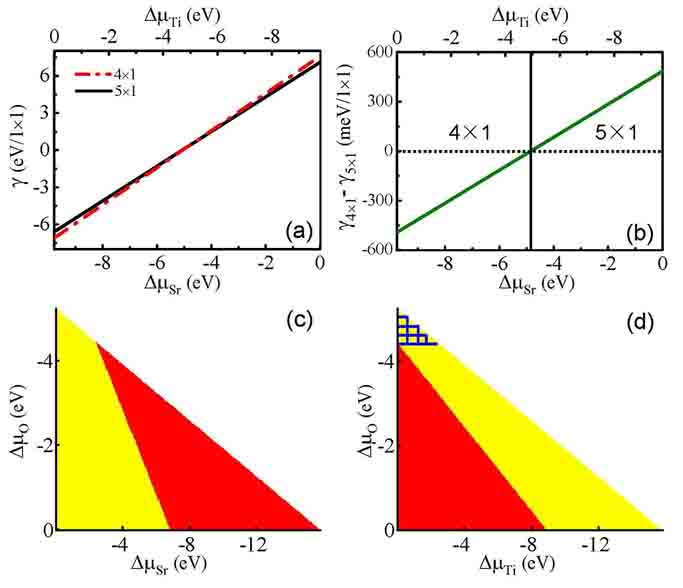}
\caption{(color online) (a) The surface free energy of (4$\times$1) (red/dashed line) and (5$\times$1) (black/solid line) as functions of $\Delta\mu_{Sr}$ and $\Delta\mu_{Ti}$, respectively ($\Delta\mu_O$ = -2 eV). (b) The energy difference of (4$\times$1) and (5$\times$1). (c) and (d) Phase stability of (4$\times$1) and (5$\times$1) upon Sr \& O and Ti \& O environments, respectively. The black (red) zones correspond the environment in which (4$\times$1) is stable while the grey (yellow) zones correspond to that (5$\times$1) is stable.}
\end{figure}

In the structural models, the nominal area concentration of Ti in (4$\times$1) is higher than that in (5$\times$1) by 0.1~ML. Experimentally (4$\times$1) or (5$\times$1) phase is formed by increasing the concentration of Ti or Sr, respectively. Quantitatively the evaporation dosage to induce the phase transition ($\sim$0.15~ML) is higher than the actual difference of the concentration due to different adhesion coefficient on surfaces of Si(100) (used for calibration) and SrTiO$_3$(110). In fact the phase stability strongly depends on the metal concentration on the surface. To further clarify, we calculate the surface free energy of (4$\times$1) and (5$\times$1) phases under different chemical environments, respectively, in terms of chemical potential. 

The surface free energy per unit area is defined by: 
\begin{equation}
\gamma=\frac{1}{2S}[G_{slab}-N_{Ti}\mu_{Ti}-N_{Sr}\mu_{Sr}-N_O\mu_O],
\end{equation}
where $G_{slab}$ is the Gibbs free energy of the slab that can be calculated with our DFT calculations, $N$ is the number of Ti, Sr, or O atoms within the slab, $\mu$ is the chemical potential of Ti, Sr, or O, respectively, while S is the area of the symmetrical surface. For the bulk SrTiO$_3$ crystal, the free energy of each unit cell is:
\begin{equation}
 G_{SrTiO_3}^{bulk}=\mu_{Sr}+\mu_{Ti}+3\mu_{O},
 \end{equation}
which equals to the energy ($E_{SrTiO_3}^{bulk}$) that can be calculated, since the surface is in the thermodynamic equilibrium with the bulk. We use relative chemical potentials $\Delta\mu_{Sr}$, $\Delta\mu_{Ti}$ and $\Delta\mu_{O}$, instead of the chemical potentials, with reference to the energies of a Ti atom in the hcp bulk structure, of a Sr atom in the cubic bulk structure and of an O atom in the O$_2$ molecule in the gas phase, respectively. And the range of accessible values for them is determined by using the method proposed by Bottin \textit{et al.} \cite{F. Bottin}. 

Figure~3~(a) shows the surface free energy as a function of $\Delta\mu_{Sr}$ or $\Delta\mu_{Ti}$ with $\Delta\mu_O=-2$~eV. In a Sr-rich ($\Delta\mu_{Sr}~\rightarrow$~0~eV, or Ti-poor, $\Delta\mu_{Ti}~\rightarrow$~-9.79~eV) environment, (5$\times$1) is more stable than (4$\times$1), while (4$\times$1) is more stable in a Sr-poor (Ti-rich) environment, with the critical concentration $\Delta\mu_{Sr}=-4.86$~eV, or $\Delta\mu_{Ti}=-4.93$~eV at $\Delta\mu_O=-2$~eV. The energy difference of the two phases shown in Fig.~3~(b) is small. This explains why it has been observed experimentally that the two phases coexist with each other or even with other (\textit{n}$\times$1) phases in fragmentary domains \cite{B.C. Russell}.

At different oxygen partial pressure, the phase transition between (4$\times$1) and (5$\times$1) can be induced by the change of Sr or Ti concentration except when $\Delta\mu_O$ is extremely low, as shown in Fig.~3~(c) and (d). At our UHV condition, we always observe the (5$\times$1) phase rather than (4$\times$1) when $\Delta\mu_{Sr}$ is increased by evaporating Sr metal onto the surface. This is in agreement with the calculation even though the phase transition is missing in an ideal O-free environment. However, the calculation predicts that the (5$\times$1) phase should be more stable than (4$\times$1) when $\Delta\mu_{Ti}$ is high and $\Delta\mu_{O}$ is low [the labeled zone in Fig.~3~(d)], totally in contradiction to the experimental observation that increasing Ti concentration induces the formation of (4$\times$1) phase in UHV. It is suggested that this is experimentally a \textquotedblleft forbidden\textquotedblright zone -- when the Ti concentration is high, the O concentration cannot be low. 

It is not surprising since DFT calculations have revealed that the TiO$_4$ tetrahedra are energetically stable and normally appear as building blocks of the reconstruction on SrTiO$_3$ surfaces, \textit{i.e.}, the surface Ti tends to be four-coordinated by oxygen on the thermodynamically stable surface \cite{J.A. Enterkin}. We calculate the formation energy of V$_O$'s on different layers, which illustrates that the energy on the top layer is more than 1~eV higher than on the buried layer. Such a characteristic is similar to that of V$_O$'s on anatase TiO$_2$(101) surface \cite{H. Cheng}. In UHV treatment, the main oxygen source is the bulk SrTiO$_3$. The undercoordinated surface Ti atoms will bind with oxygen atoms diffusing from bulk and form the TiO$_4$ tetrahedra. Kinetically oxygen is very diffusive relative to cations in bulk SrTiO$_3$ \cite{A.E. Paladino}. In Fig.~4 (a)-(c), we schematically show how an V$_O$ diffuses from the (110) surface towards the bulk. Associated with two 3-coordinated Ti atoms, a surface V$_O$ is unstable [Fig.~4~(a)]. It diffuses to the second layer and then to the fourth layer [Fig.~4~(b) and (c)]. The energy barrier for both processes is in the order of 0.6~eV as calculated by Cuong \textit{et al.} \cite{D.D. Cuong}. Such barriers can be overcome during annealing at $1000\,^{\circ}\mathrm{C}$ and the equilibrium is reached with V$_O$'s diffusing deep into the bulk.

Comparing with the theoretical simulations [Fig.~4~(e)-(h)], the bright spots labeled with circles in the STM image [Fig.~4~(d)] can be attributed to V$_O$'s or hydroxyl-adsorbed V$_O$'s (reacted with residual water) on the surface. Statistics of a number of images show that the possible V$_O$'s density is $\leq$~0.4$\%$~ML, where one ML is defined as the number of oxygen on a perfect (4$\times$1) surface. It is suggested that the thermodynamically stable surface is experimentally reached. 

\begin{figure}
\includegraphics[clip,width=3.4in]{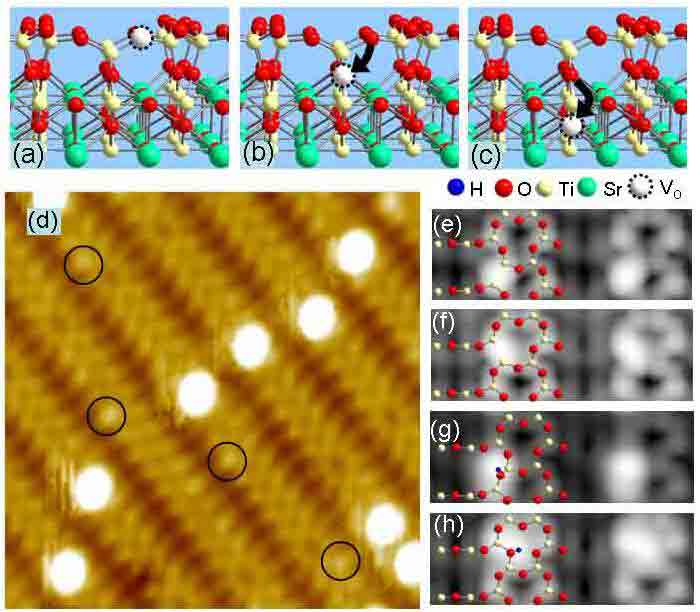}
\caption{(color online) (a) An STM image (10$\times$10~nm$^2$, 2.0~V) of (4$\times$1) on SrTiO$_3$(110). The circles indicate possible V$_O$'s or hydroxyl-adsorbed V$_O$'s. (b)-(e) The simulated STM images of (4$\times$1) surface with an oxygen vacancy or a hydroxyl-adsorbed V$_O$ on different sites. (f)-(h) A schematic diagram of the pathway of an V$_O$ diffusing from SrTiO$_3$(110) surface towards the bulk.}
\end{figure}

It has been a great challenge to control the V$_O$'s density or to minimize it to obtain an oxide surface or other LD structures with ideal stoichiometry. Highly reactive oxygen sources are employed at a high partial pressure during the growth with molecule beam epitaxy (MBE) or pulsed-laser deposition (PLD) technique. However, such a tuning is limited since the growth mode of films changes from two-dimensional layer-by-layer to three-dimensional island growth when the oxygen pressure exceeds 10$^{-2}$ mbar \cite{M. Huijben}. Moreover, specifically in MBE, the oxidation of the metal sources is also an important issue. The oxygen properties near SrTiO$_3$(110 ) surface region suggests a method to epitaxially grow high-quality titanate films along [110] -- as long as the metal stoichiometry is well controlled, V$_O$'s can be eliminated with low growth rate at high temperature.

In conclusion, we obtain atomically well defined (4$\times$1) and (5$\times$1) reconstructions on SrTiO$_3$(110) surface. With \textit{ab initio} calculations we resolve the microscopic structures and explain the phase transition between the two phases in terms of the surface free energy determined by metal concentration. Moreover, the V$_O$'s density on the reconstructed SrTiO$_{3}$(110) surface could be extremely low, suggesting a feasible way to grow high-quality thin films of perovskite titanates.

\begin{acknowledgments}
The authors are grateful for the discussion with Jiandi Zhang (LSU). This work is supported by NSFC (10704084 \& 11074287) and MOSTC (2007CB936800). The calculation results are obtained on the Deepcomp7000 of Supercomputing Center, Computer Network Information Center of Chinese Academy of Sciences.
\end{acknowledgments}

\end{document}